\documentclass[conference]{IEEEtran}
\IEEEoverridecommandlockouts
\usepackage{cite}
\usepackage{amsmath,amssymb,amsfonts}
\usepackage{booktabs}
\usepackage{algorithmic}
\usepackage{graphicx}
\usepackage{textcomp}
\usepackage{xcolor}
\usepackage{tipa}
\usepackage{url}
\def\BibTeX{{\rm B\kern-.05em{\sc i\kern-.025em b}\kern-.08em
    T\kern-.1667em\lower.7ex\hbox{E}\kern-.125emX}}
\begin{document}

% % \title{Why Big Techs TTS fail to reproduce Brazilian Accents}
% \title{Exposing the Phonetic ``uncanny valley'' of Brazilian Portuguese TTS}
% \title{All at Once: Exposing the Phonetic "Uncanny Valley" of Brazilian Portuguese TTS}
% \title{Accent Dilution as a Feature: Detecting Synthetic Brazilian Portuguese Speech with Phonological Markers}
\title{Synthetic speech detection in Brazilian Portuguese through accent-related features}

\author{\IEEEauthorblockN{Pedro H. L. Leite}
\IEEEauthorblockA{\textit{PEE/COPPE} \\
\textit{UFRJ}\\
Rio de Janeiro-RJ \\
pedro.lopes@smt.ufrj.br}
\and
\IEEEauthorblockN{ Pedro Benevenuto Valadares}
\IEEEauthorblockA{\textit{FEEC} \\
\textit{UNICAMP}\\
Campinas-SP \\
p204483@dac.unicamp.br}
\and
\IEEEauthorblockN{Luiz W. P. Biscainho}
\IEEEauthorblockA{\textit{DEL/Poli \& PEE/COPPE} \\
\textit{UFRJ}\\
Rio de Janeiro-RJ \\
wagner@smt.ufrj.br}
}

\maketitle

\begin{abstract}
Leading commercial and open-source Text-to-Speech (TTS) models fail to emulate the regional phonetic diversity of Brazilian Portuguese (pt-BR). By aggregating disparate dialects into a single training distribution, they generate a synthetic ``diluted'' accent: a phonetic profile attempting to represent all regional distributions simultaneously, but ultimately carrying phonological ambiguity dissociated from natural socio-phonetic realizations. This work introduces a speech deepfake detection methodology combining multilingual phone recognizers with classical signal processing to extract phoneme-level features in consonantal and vocalic realizations with high geographic variance. The analysis reveals that the distributional gap over these features suffices to distinguish natural and synthetic voices through unsupervised Kernel Density Estimation, establishing dialectal inconsistency as a useful and interpretable feature for spoofing detection in pt-BR. Evaluation on pt-BR anti-spoofing datasets shows that these explainable, lightweight, low-dimensional features can boost the performance of foundation models on the task, and show generalization capabilities in a cross-dataset leave-one-out setup.
\end{abstract}

\begin{IEEEkeywords}
Brazilian Portuguese, text-to-speech, regional accent variation, phonetic feature extraction, spoofing detection, speech synthesis
\end{IEEEkeywords}

\section{Introduction}

The adoption of synthetic speech in voice agents, virtual assistants, content creation, and accessibility tools is turning speech synthesis, and, more specifically, Text-to-Speech (TTS) into a standard component of modern human-agent interaction. Despite recent significant progress, regional accent features remain under-addressed by such systems. It has been shown~\cite{michel2025itsnotme} that dialects outside of mainstream English speaking countries are often misrepresented, which has a profound impact on how human listeners perceive and trust synthetic speech content.

% On the other hand, modern commercial models are frequently trained on massive datasets[ref] that, due to scalability limitations, make it impossible to label and curate accent diversities. This introduces a gap in the quality of the systems, since they fail to account for regional diversity by collapsing disparate dialects into a single training distribution~\cite{xinyuan2025scalable}.
% This problem is even greater with languages outside of Mandarin and English, which have received less research attention[ref].

This issue is largely driven by the scaling paradigms of modern commercial TTS architectures. To achieve broad linguistic coverage, they are typically trained on massive
%, softly curated and 
multilingual datasets whose scale make manual labeling and curation of regional accents functionally impossible. This introduces a gap in accounting for regional diversity, since disparate dialects are collapsed into a single training distribution~\cite{xinyuan2025scalable}. 

In languages outside the scope of highly researched ones like English and Mandarin, this gap is even more prominent: for example, in the specific case of Brazilian Portuguese (pt-BR), no large-scale, accent-annotated dataset currently exists. This data scarcity often prompts developers to treat highly distinct regional variants as a single distribution, or worse, to aggregate disparate cross-continental variants like European Portuguese (pt-PT) into unified corpora, severely eroding phonetic specificity~\cite{matos2024accent}. Furthermore, as these languages are generally trained against a multilingual objective together with other languages and almost never alone, it is likely that in-language phones are contaminated by foreign realizations.

% In this context, Brazilian Portuguese (pt-BR) does not have any large-scale accent-annotated dataset yet[ref], while having many distinguishable accent groups being threated as one by modern speech synthesizers. Furthermore, the aggregation of very distinct linguistic variants such as European Portuguese (pt-PT) into unified corpora results in an even worse loss of phonetic specificity~\cite{matos2023accent}.

% In agreement with the findings of~\cite{itsnotarepresentation}, according to~\cite{lee2026exploring}, accurate accent representation is a matter of social identity but also cognitive processing. 

Beyond the cultural identity issue mentioned in~\cite{michel2025itsnotme}, such dialectal imprecision carries significant cognitive and experiential costs. Linguistic features are central to how users perceive the personality, reliability, and fluidity of a voice assistant, and that affects the general preference for voice agents~\cite{lee2026exploring}. When synthetic voices fail to maintain a consistent regional identity, they can create ``speaker drift'' effects, which have been shown to sabotage the coherence of speech~\cite{huang2026speakerdrift} and disrupt conversational experience. Furthermore, perceptual inconsistency has cognitive costs: listeners (and machines) process speech more efficiently when phonemes are consistent.

While there is a clear demand for TTS with more accurate accents~\cite{zhong2025pairwise}, technical implementation is still considerably challenging. Accent variants involve both phonetic and prosodic components~\cite{liu2022controllable} that are often targeted/modeled by large self-supervised learning (SSL) models such as HuBERT~\cite{hsu2021HuBERT} and Wav2Vec~\cite{schneider2019wav2vec}. Yet, these ``black box'' architectures might fail to capture sociophonetic diversity effectively. For instance, in the specific case of Brazilian Portuguese, these models still struggle with cross-domain accent classification~\cite{matos2024accent}.

On the other hand, the rapid advancement of speech synthesis has surfaced novel critical security vulnerabilities. With the rise of audio deepfakes, social engineering attacks, financial fraud, and bypassing automatic speaker verification (ASV) systems, security has become a real challenge in many sectors. Although benchmark initiatives such as ASVspoof challenges have driven significant progress in countermeasures, recent evaluations highlight a significant generalization gap across unseen generative models ~\cite{muller2022doesgeneralize}. Consequently, there is a pressing need within the anti-spoofing community to identify different methodologies for synthesis detection that remain persistent even for state-of-the-art synthesizers and that do not overfit on acoustic cues or superficial dataset features.

Having that in mind, this work proposes an alternative angle to deepfake detection that relies on measuring accent realization consistency. It builds directly on the signal processing pipeline specifically introduced by~\cite{leite2026extracting} for pt-BR, but shifted to a different question: is it possible to operationalize the phonetic inconsistency in TTS training pipelines as an acoustic clue for anti-spoofing? Just as early generative image models struggled to render human hands, leaving telltaling artifacts in the generated images, the aim here is to show that commercial and open-source TTS models still leave many detectable artifacts in the sociophonetic domain of pt-BR.

The main contributions of this article are then the following:
\begin{itemize}
    \item A comparative analysis of phonetic distributions across major commercial and open source TTS engines over a curated dataset is provided. The evaluation demonstrates that while these models are widely deployed, they exhibit many phonetic inconsistencies that can help distinguish them from natural speech in general.
    Experimental results show that these ``dialectal anomalies'' allow for the detection of synthetic speech with high accuracy as a separable feature space, even without supervised classification.
    
    \item A thorough analysis is conducted regarding feature interpretability, showing from different angles where modern TTS may fail to reproduce real accent realizations.
    
    \item A comprehensive benchmark of the proposed methodology against traditional accent classification and anti-spoofing datasets is presented, providing a quantifiable way to assess the practical usefulness of sociophonetic features in detecting speech synthesis.
\end{itemize}

\section{Related Works}
\label{sec:related-works}

Starting with data availability, the ASVspoof challenge series~\cite{wang2020asvspoof2019, yamagishi2021asvspoof2021, wang2026asvspoof5} produced many datasets for anti-spoofing in English, while WaveFake~\cite{frank2021wavefake} also included Japanese samples with different vocoder architectures instead of end-to-end generated samples. Furthermore,~\cite{yamagishi2021asvspoof2021} introduces ``in-the-wild'' data to test for detection generalization over real-world scenarios. Recently, MLAAD~\cite{muller2024mlaad} has been established as the main multilingual reference dataset for synthesis detection, and includes speech in Brazilian Portuguese.

Specifically for pt-BR, BRSpeechDF~\cite{filho-etal-2025-brspeech} stands out as the main large-scale language-specific dataset available; and the FakeBR Accent corpus~\cite{fakebr_accent2025} should also be mentioned, as it contains synthetic samples of various pt-BR accents.

While not explicitly designed for anti-spoofing or accent modeling, datasets containing a variety of speech from many regions of Brazil are also highly relevant. Notable examples include CORAA~\cite{candido2023coraa}, Mozilla Common Voice~\cite{ardila2020common}, ColingPB~\cite{stein2015colingpb}, CML-TTS~\cite{oliveira2023cmltts}, NURC-SP~\cite{rodrigues2024nurcsp}, NURC-RE~\cite{oliveirajr2016nurcdigital}, Certas Palavras~\cite{araujo2026certas}, CETUC~\cite{alencar2008lsf}, and Tagarela~\cite{deoliveira2026tagarela}. Although these datasets are diverse in accent realizations, none of them provide reliable, large-scale accent annotations.

Given the scarcity of labeled data (a problem not unique to Brazilian Portuguese), prior works have pursued alternative paths that do not necessarily rely on sociophonetic labels. While \cite{xinyuan2025scalable} proposes a geo-location predictor for a soft supervision approach, \cite{lertpetchpun2026quantifying} and~\cite{leite2026extracting} employ phonological rule detection to classify accents without predefined labels.

Another branch of resources useful to the accent identification pipeline are multilingual phonetic transcribers (phone recognizers). These systems learn to map acoustic features into phonetically unambiguous characters such as the IPA~\cite{ipa1999handbook}. Recent literature relies on deep learning-based approaches:  Allosaurus~\cite{li2020allosaurus} and Wav2Vec2Phoneme~\cite{xu2022wav2vec2phoneme} appeared first, followed recently by ZIPA~\cite{zhu2025ZIPA}, CUPE~\cite{rehman2025cupe} and PhoneticXeus (PX)~\cite{bharadwaj2026phoneticxeus}, with more robust training and data pipelines.

Using phonetic features directly for deepfake detection,~\cite{yang2026forensic} analyzes segmented vowel formant distributions, while~\cite{zhu2024slim} proposes a framework to find a style-linguistic mismatch in speech, and~\cite{phonemedf2026} extracts segmented phoneme features using SSL models to build an English phoneme-level deepfake detection dataset. Also relevant to this context,~\cite{warren2025pitch} focuses on acoustic prosodic modeling for anti-spoofing.

Building on the many English/Mandarin and multi-lingual datasets and challenges,~\cite{li2025where} has shown that significant performance can be attained over many languages and systems with foundation models such as Wav2Vec2Bert~\cite{seamless2023multi}, but it also highlights that commercial TTS detection still represents a reasonable gap even for them. 

Furthermore,~\cite{muller2022doesgeneralize} has shown that models trained on ASVspoof challenge data can suffer severe performance drops when applied to out-of-domain data. This is also highlighted in the BRSpeechDF report, where the pt-BR evaluation ended up with higher Equal Error Rate (EER) and lower accuracy than ASVSpoof English baselines using the same models.

In summary, foundation models trained for anti-spoofing in English struggle to adapt to pt-BR anti-spoofing datasets and commercially competitive TTS systems, while phonological feature extraction gains traction in the literature. The lack of reliably annotated data also hinders the development of foundation models for language-specific and accent-related features in Brazilian Portuguese. This further motivates the alternative pipeline proposed here, which relies on the use of lightweight, localized, and acoustically extractable phonological features applied to the task of synthetic speech detection.

\section{Methodology}
\label{sec:methodology}
The methodology for studying the impact of accent features on the naturalness of synthetic speech is described below. 

\subsection{Accent varieties dataset}
\label{subsec:exp1}

The established paradigm to compare feature performance in classification is to train and test on extensive cross-domain anti-spoofing data. However, the only large-scale Brazilian Portuguese anti-spoofing training dataset is the recent BRSpeechDF. FakeBR Accent would also be an option, but it is reduced in size and contains only one TTS provider option. As this data source scarcity raises the likelihood of overfitting into channel/domain specific learning that does not generalize, both datasets should be more useful for comparison with previous literature benchmarks than for real generalization tests.

With that in mind, a curated custom dataset was built, containing a sufficient amount of speech of both natural and synthetic speakers, specifically designed to support accent diversities on the natural side, and to avoid adaptation issues outside the training domain. The intuition is to evaluate models using cross-validation with leave-one-system-out procedures, in order to quantify the out-of-domain performance without losing natural speaker references, and also to monitor the false positives, i.e., naturals being detected as synthetic speech.

The natural side comprises 364 speakers drawn from many different corpora~\cite{candido2023coraa, ardila2020common, stein2015colingpb,oliveira2023cmltts, rodrigues2024nurcsp, oliveirajr2016nurcdigital,araujo2026certas, alencar2008lsf, deoliveira2026tagarela},  chosen in such a way that their distribution covers multiple dialects and multiple recording chains, preserving sample quality. The synthetic side comprises 57 TTS voices across eight providers, four commercial --- Azure~\cite{azure2024tts}, Google~\cite{google2024tts}, OpenAI~\cite{openai2024tts}, and ElevenLabs~\cite{elevenlabs2024tts} ---
and four open-source --- F5-TTS~\cite{chen2025f5tts},
Qwen3-TTS~\cite{hu2025qwen3tts}, Piper~\cite{piper2023}, and Kokoro~\cite{kokoro2024}. All models synthesized the same $50$ sentences, carefully designed for phonological completeness and diversity using balanced text subsets coming from the natural data, as a best effort to match the natural side\footnote{Recipes for downloading data, synthesizing speech and additional content about the experiments are available at \url{https://gpa-smt-ufrj.github.io/slt2026/}}.

\subsection{Accent-related feature extraction}
\label{subsec:feature-extraction}
Building upon the localization pipelines proposed in~\cite{leite2026extracting} and~\cite{yang2026forensic}, the core acoustic feature extraction block also relies on localizing accent-related phones.
To this end, ZIPA is used as timestamp extractor/aligner and to provide phone realization logits, along with PhoneticXeus for an ensemble classification. 

Aside from consonant disambiguation, formant space for aligned vowels is also evaluated. In Brazilian Portuguese, /o/ and /e/ openness are also tied to regionalistic realizations, as found in ALiB maps~\cite{cardoso2015atlas}. Furthermore, the study looks for vowel contaminations that might come from another language (i.e. occupation of non-natural vowel space for Brazilian Portuguese speakers). The intuition behind extracting features at such ambiguous places (in both consonants and vowels) is that there is much variance over natural speech there, and models might smoothen the distribution on the presence of variance or choose wrong realizations that do not sound natural.

Since accent is intrinsically a speaker-level trait rather than a property of an isolated audio clip, the methodology proposed here diverges from standard literature, which predominantly performs deepfake detection strictly at the utterance level. Instead,  the extracted phonological features are aggregated across multiple utterances to construct a comprehensive phonetic profile for each speaker. While this requires a larger number of samples per speaker to build a robust profile, it can bring explainability, easier generalization and overall robustness in detection against noisy, superficial acoustic features that can confuse utterance-level detectors.

This way, the phonological analysis relies on a combination of aggregated vectors per speaker. The consonant realization features are the 40-dimensional ZIPA logits vector for the aggregated possible consonant realizations across the three extraction tasks concatenated with 6 spectral moments as defined in~\cite{leite2026extracting}, with the addition of the 15-dimensional PhoneticXeus prediction probabilities obtained through a similar extraction procedure. For vowels, the parselmouth~\cite{parselmouth} python package is used to extract vowel formants and pitch, building a 21-D vector of mean F1/F2/F3 vectors for the 7 standard vowels of the language (\textipa{a, E, e, i, O, o, u}), alongside 6-D LTFD and 2-D LTF0 as defined in~\cite{yang2026forensic}.

Embeddings from foundation models are also extracted at such critical accent-related acoustic moments, increasing feature retrieval power, but possibly losing interpretability. These embeddings will also serve as a baseline for feature effectiveness in the experimental tasks.

To ensure that the detection models isolate the targeted linguistic phenomena rather than overfitting to superficial acoustic artifacts or channel biases, all natural and synthetic audio samples were loudness-normalized and standardized to a 16-kHz sampling rate before each feature extraction.

% To quantify this aggregated profile, we define a family of speaker-level naturalness metrics, expanding prior methods presented on  ~\cite{leite2026extractingaccentfeaturesspoken}, to explain the classifier's decisions:
% \begin{itemize}
%     \item \textbf{ADS (Accent Deviation Score):} Measures the distance to the nearest natural phonological class.
%     \item \textbf{VNS (Variance Naturalness Score):} Evaluates whether the within-speaker variance falls inside the natural distribution band.
%     \item \textbf{JMD (Joint Mahalanobis Distance) and CDC (Cross-Dimensional Coherence):} Assess joint profile coherence to detect impossible combinations of phonetic markers.
% \end{itemize}

\section{Experiments}
\label{sec:experimental-design}
The experimental framework is divided into two distinct groups, to evaluate both the fundamental separability of the proposed features and their practical utility across different data paradigms and performance measurements against established spoofing detection baselines in Brazilian Portuguese.
\begin{figure}[h]
\includegraphics[width=\columnwidth]{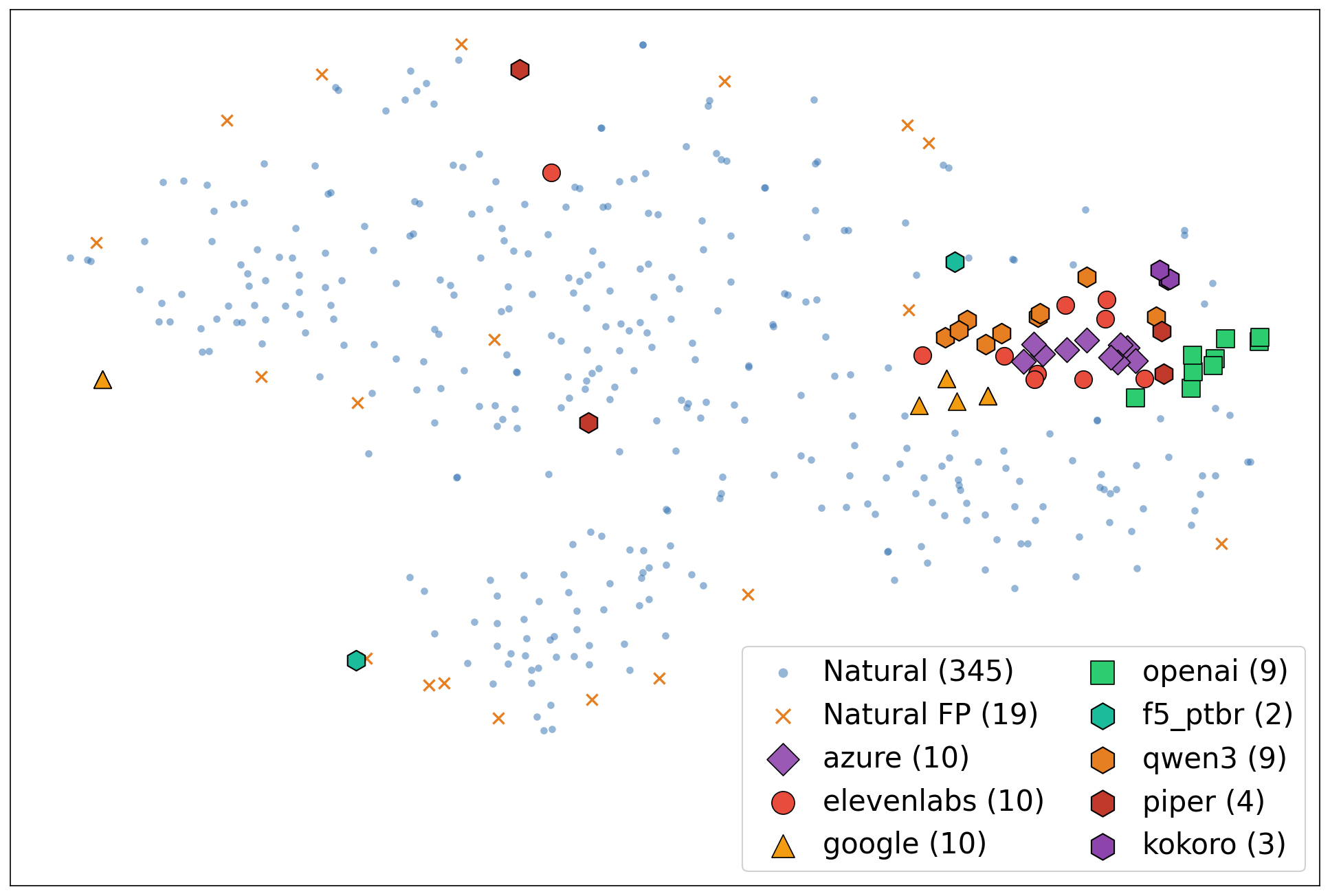}
\caption{t-SNE plot for phonological features. Labels show the system and corresponding number of speakers. TTS systems and speakers are grouped into a single cluster or at the boundaries of the natural distribution.}
\label{fig:tsne}
\end{figure}

\subsection{Explainable features}
\label{subsec:explainable-feat}
The first set of experiments sought to explore the distribution of phonological features in synthetic and natural samples. The consonant and vowel vector distributions are first plotted under the t-SNE~\cite{vandermaaten2008visualizing} dimensionality reduction technique, as shown in Figure~\ref{fig:tsne}. To analyze the distribution of natural features, a Kernel Density Estimator (KDE)~\cite{silverman1986density} is fitted on natural samples only. This detects 100\% of synthetic samples as being out-of-distribution (OOD) without leaving any of the naturals out, but to better visualize the boundaries, in the illustration the threshold has been raised to intentionally leave 5\% of naturals as OOD.

Although the plot and KDE analysis already indicate reasonable separation between classes, looking at true negative rates is also needed, since non-fit natural samples might still fall into OOD. Therefore, two follow-up experiments are presented to further validate the separation: i) low-dimensionality cross-validated KDE classification using PCA~\cite{jolliffe2002principal} and ISOMAP~\cite{tenenbaum2000global}; ii) cross validation across supervised classification algorithms (XGBoost~\cite{chen2016xgboost}, Random Forest~\cite{randomforest}, SVM~\cite{Cortes1995} and logistic regression~\cite{cox1958logregression}) using the accent-related features. Experiment ii) also includes combinations and performance comparisons with general-purpose foundation models (Wav2Vec2Bert~\cite{seamless2023multi}, XLSR~\cite{conneau2021unsupervised}, ECAPA-TDNN~\cite{desplanques2020ECAPA}, and Hubert~\cite{hsu2021HuBERT}).

\begin{table}[h]
    \centering
    \caption{Unsupervised KDE detection on phonological features (75D), AUC.
    $^{\dagger}$t-SNE is transductive (no out-of-sample adaptation; not deployable).}
    \label{tab:auroc}
    \begin{tabular}{lc}
      \toprule
      Method & AUROC (\%) \\
      \midrule
      t-SNE (2D)$^{\dagger}$ & 93.9 \\
      Isomap (15D)           & 74.3 \\
      PCA (5D)               & 72.1 \\
      \bottomrule
    \end{tabular}
\end{table}
Table~\ref{tab:auroc} shows the unsupervised KDE classification Area under ROC over the 5-fold system held out procedure: low dimensionality unsupervised clustering already shows that separation is still possible without losing too many true negatives.  

\begin{table}[!h]
\centering
\caption{Supervised detection, mean $\pm$ SD over cross-validation (CV). }
\label{tab:exp43_clf}
\begin{tabular}{llrcc}
\toprule
Feature set & Classifier & Dim & ACC (\%) & EER (\%) \\
\midrule
ZIPA+XLS-R & LogReg & 1070 & 97.1 $\pm$ 3.7 & 1.5 $\pm$ 3.2 \\
XLS-R & LogReg & 1024 & 96.8 $\pm$ 3.7 & 1.8 $\pm$ 3.7 \\
ZIPA+ECAPA & RandomForest & 238 & 96.7 $\pm$ 3.5 & 2.2 $\pm$ 3.8 \\
ZIPA+HuBERT & LogReg & 1070 & 96.5 $\pm$ 4.0 & 2.5 $\pm$ 4.3 \\
ZIPA+PX & RandomForest & 61 & 96.2 $\pm$ 4.2 & 2.5 $\pm$ 4.0 \\
ZIPA+Form.+PX & RandomForest & 90 & 95.9 $\pm$ 4.7 & 3.3 $\pm$ 4.5 \\
ZIPA+W2V2B. & LogReg & 1070 & 95.6 $\pm$ 4.3 & 3.3 $\pm$ 4.6 \\
ZIPA & RandomForest & 46 & 95.7 $\pm$ 4.5 & 3.3 $\pm$ 4.9 \\
HuBERT & LogReg & 1024 & 95.4 $\pm$ 4.7 & 3.4 $\pm$ 5.2 \\
ZIPA+Formants & XGBoost & 75 & 94.9 $\pm$ 5.3 & 3.9 $\pm$ 5.5 \\
W2V2B. & LogReg & 1024 & 95.2 $\pm$ 4.5 & 3.9 $\pm$ 5.0 \\
PX & SVM-RBF & 15 & 90.4 $\pm$ 5.7 & 9.8 $\pm$ 6.2 \\
Formants+PX & RandomForest & 44 & 89.0 $\pm$ 6.8 & 10.2 $\pm$ 6.7 \\
ECAPA & SVM-RBF & 192 & 88.5 $\pm$ 6.4 & 10.4 $\pm$ 6.5 \\
Formants & SVM-RBF & 29 & 82.8 $\pm$ 7.4 & 16.7 $\pm$ 7.4 \\
\bottomrule
\end{tabular}
\end{table}
The ablation results for feature combination and comparison with foundation models under cross validated evaluation are shown in Table~\ref{tab:exp43_clf}. It can be noticed that many combinations of phonological features alone have classification capabilities similar to that of SoTA foundation models. Also, many of the best classifiers combine these models with the accent-related feature space, indicating that the proposed pipelines should provide additional cues towards better classification.

\begin{figure}[h]
\includegraphics[width=\columnwidth]{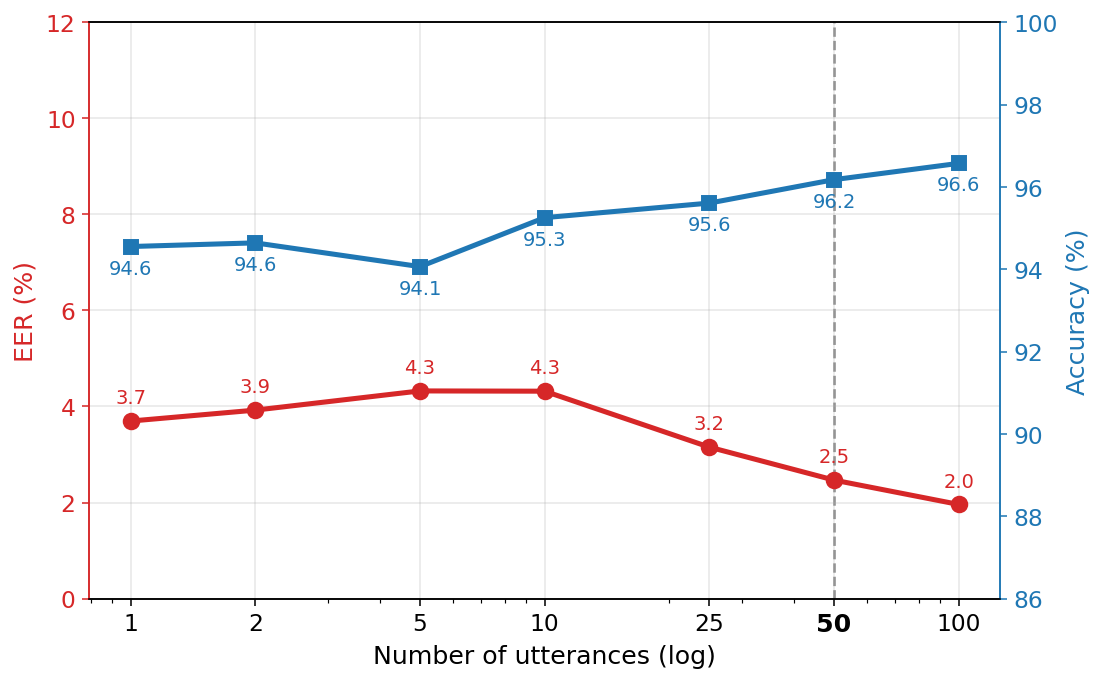}
\caption{Accuracy and detection rates against the number of utterances per speaker.}
\label{fig:budget}
\end{figure}
To illustrate the impact of utterance count on detection rates, the dataset was expanded to 100 utterances per speaker, measuring scaling from 1 to 100. Figure~\ref{fig:budget} reports detection rates for the best speaker-aggregated supervised model, showing stability below 10 utt/speaker and near log-linear growth (or error decrease) above it. This aligns with models actually learning the consistency gaps, which emerge only once a minimum number of utterances is available and strengthen as that number grows. The experiments were capped at 50 samples/speaker as a reasonable data-accuracy trade-off.

\begin{figure*}[t]
    \centering
    \includegraphics[width=\textwidth]{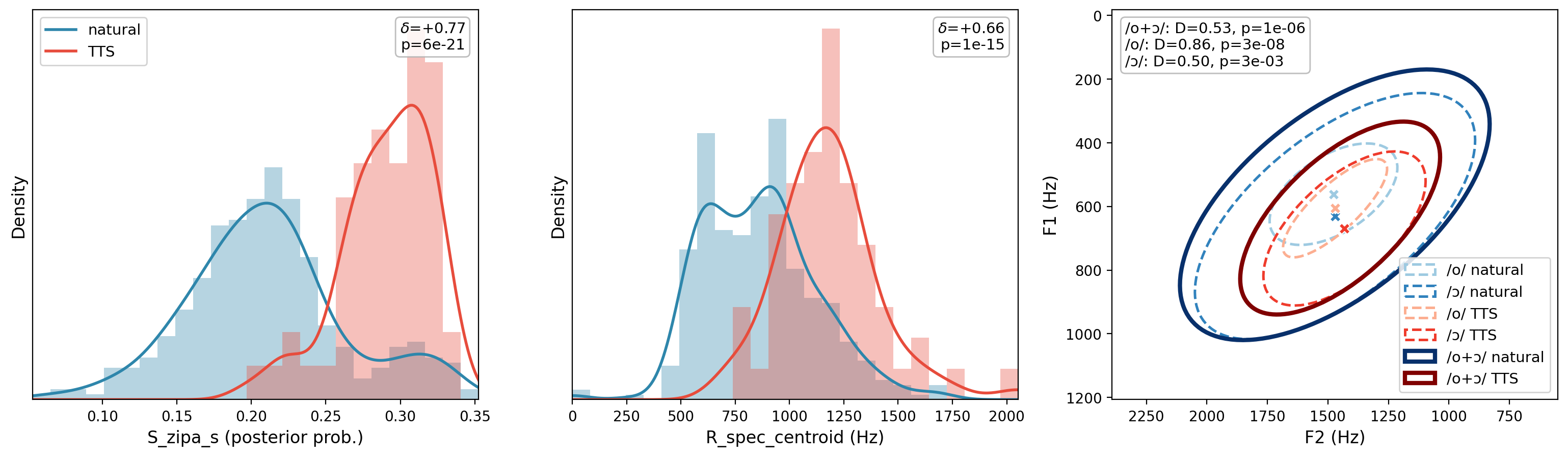}
    \caption{Per-speaker distributions of some of the most discriminative interpretable features for
    natural vs.\ synthetic (TTS) Brazilian Portuguese speech. Each panel quantifies the natural--synthetic difference with an effect size and a corresponding $p$-value (Cliff's $\delta$~\cite{cliff1993dominance} with Mann-Whitney U~\cite{mann1947on}, or Mahalanobis
distance~\cite{mahalanobis1936generalised} with Hotelling's $T^2$~\cite{hotelling1931generalization}
for the vowel panel). }
    \label{fig:exp43_triptych}
  \end{figure*}

\subsection{Interpretability analysis}
\label{subsec:interpretability}
With the proposed phonological procedures, we can directly look at feature distributions to understand what kind of signals the models are learning to distinguish. 

In Figure~\ref{fig:exp43_triptych}, we show the natural against synthetic speech comparison for three feature distributions\footnote{More feature distributions are available at the companion website.}. The first panel (on the left) corresponds to the sibilant /s/ probability distribution for ZIPA prediction when /s/-coda happens. It is possible to observe that there is much synthetic activity where there is not for natural speech. In the middle panel, the centroid of energy distribution (one of the 6 spectral moments retrieved) is shown for the /r/-coda task: synthetic shape differs from natural occurrences, where human speakers present some multimodality and a longer tail for higher frequencies. The last of the three panels illustrates the two first formant distributions for open and closed /o/ vowels (\textipa{O} and \textipa{o}), where ellipses show where 95\% percent of the points fall for each split. It is possible to infer that natural speakers cover a broader space, and synthetic samples are slightly biased.  It is important to note that the reported $p$-values are exploratory and descriptive of the selected features used to illustrate separability. In the three cases (and also in many other features), the statistical tests point to significant distributional differences.

These observations corroborate the hypothesis that synthetic speech systems have difficulty dealing with regional ambiguities, since the analysis of these high-variance phone realizations reveals different distribution shapes and localizations.

Alongside the distribution mismatches, another experiment is setup for measuring accent-related consistency. Especially for consonants, it is expected that within a single enunciation, the same speaker maintains reasonable consistency over phone realization choices. To measure this, phonetic realization classifiers similar to the ones described in~\cite{leite2026extracting} are trained in subsets of our customized dataset for the three ambiguous consonants, holding out training datasets from inference data to avoid speaker and channel leakage. Next, Shannon's bit entropy is adopted to quantitatively compare the switching between classes for a single utterance:
\begin{equation*}
  H(p) = -\sum_{k=1}^{N} p_k \log_2 p_k,
\end{equation*}
where $p_k$ is the probability of a class given a reference realization (ex. /s/-coda sounding as ``sh''). This way, higher entropy values mean more switching and less consistent realizations. For this experiment, signals with less than two ZIPA detections per marker have been filtered out, which led to the need for longer sentences and also required the synthesis of additional longer texts to match the number of markers per sentence of naturals. This resulted in an average of about three detections per signal for each marker, and around 1000 utterances for both synthetic and natural speakers\footnote{Classifier accuracies for this experiment are reported in the companion page.}. 

\begin{figure}[!h]
    \centering
    \includegraphics[width=0.9\columnwidth]{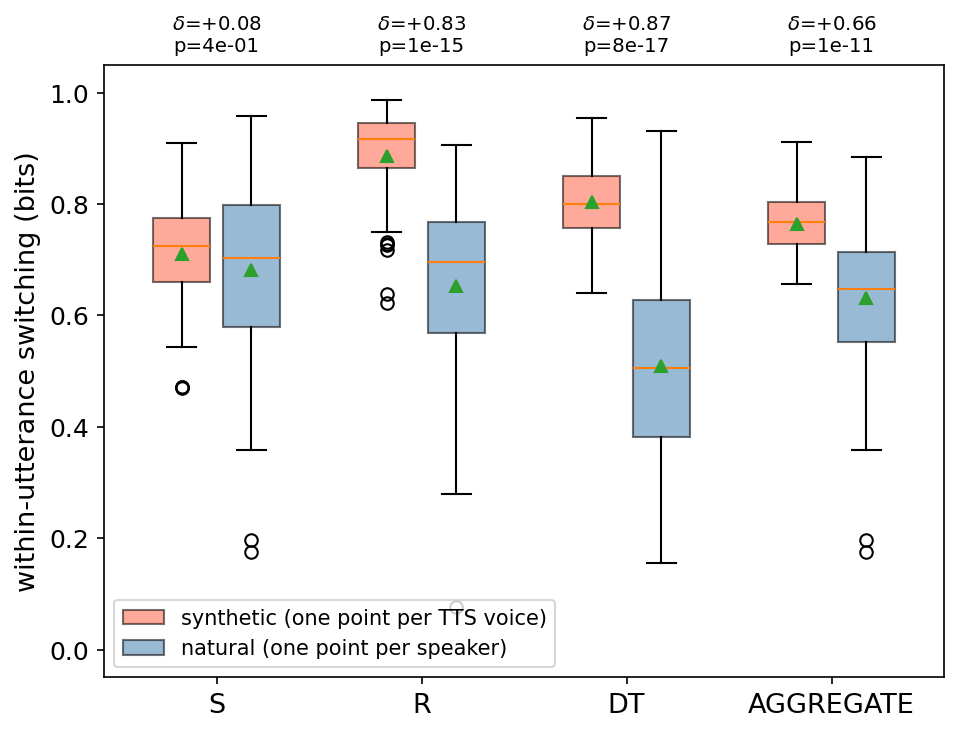}
     \caption{Within-utterance marker switching, averaged per
  voice/speaker.}
    \label{fig:exp39_permarker}
\end{figure}
Since utterances are repeated measures of the same voices and speakers. $H(p)$ is averaged within each unit (TTS or natural speaker). Figure~\ref{fig:exp39_permarker} shows the result: synthetic speech switches more for /r/-coda ($\delta=+0.83$) and /d,t/-palatalization ($\delta=+0.87$), and in aggregate ($\delta=+0.66$) --- all with $p<10^{-10}$ ---, while /s/-coda shows no significant difference. This aligns with the hypothesis that synthetic models dilute regional accents into less consistent realization spaces.

% Exp 43d -- best classifier ALGORITHM per feature set, TTS-vs-natural (BP KDE pool, 364 natural / 57 synthetic; NO MLAAD). Generated from exp43_clf_algo_results.csv. Requires \usepackage{booktabs}. Balanced 5-fold CV (20 seeds x 5 folds), StandardScaler; mean $\pm$ SD. For each feature set the algorithm with the lowest EER among {LogReg, SVM-RBF, RandomForest, MLP, XGBoost} is reported. Rows sorted by EER (ascending).

\subsection{Deepfake Detection in Consolidated Benchmarks}
\label{subsec:exp2}
The second group of experiments evaluates the utility of the proposed features on consolidated antispoofing datasets for Brazilian Portuguese, more specifically BRSpeechDF, FakeBRAccent and the Portuguese subset of MLAAD.

%For the standardized evaluations, we extracted features, trained and tested exclusively on each one, FakeBRaccent or BRSpeech-DF. 
Starting with BRSpeechDF, the original splits are used to train, validate, and test models. Based on the results presented in Table~\ref{tab:brsdf_indomain}, it can be seen that the baseline is almost matched by ZIPA features alone, although real gain is observed when training large models on the critical moments. Furthermore, adding interpretable features generally helps, providing an additional advantage to the best combination of features. 
\begin{table}[h]
  \centering
  \caption{In-domain deepfake detection on BRSpeech-DF (official splits)}
  \label{tab:brsdf_indomain}
  \begin{tabular}{lrcc}
  \toprule
  Feature set & Dim & ACC (\%) & Test EER (\%) \\
  \midrule
  \textbf{ECAPA+W2V2BERT+ZIPA} & \textbf{1262} & \textbf{97.14} & \textbf{2.00} \\
  ECAPA+W2V2BERT & 1216 & 97.15 & 2.06 \\
  W2V2BERT & 1024 & 96.91 & 2.27 \\
  ZIPA+W2V2BERT & 1070 & 96.95 & 2.47 \\
  ECAPA+ZIPA & 238 & 90.65 & 11.92 \\
  ECAPA & 192 & 90.43 & 12.12 \\
  XLS-R & 1024 & 80.51 & 26.52 \\
  HuBERT & 1024 & 73.93 & 31.82 \\
  ZIPA & 46 & 72.27 & 37.27 \\
  \midrule
  SLS-ECAPA (baseline)~\cite{filho-etal-2025-brspeech} & 192 & 75.97 & 30.67 \\
  \bottomrule
  \end{tabular}
\end{table}

% Exp 43d -- supervised classifier ablation, TTS-vs-natural over the BP KDE pool (364 natural / 57 synthetic speakers, NO MLAAD). Generated from exp43_clf_results.csv. Requires \usepackage{booktabs}. Balanced 5-fold CV (20 seeds x 5 folds, natural subsampled to synth size), XGBoost; mean $\pm$ SD. clf7/clf7px/switch use leakage-free speaker-grouped GroupKFold cross-fitting. Rows sorted by AUC.
% Exp 43d -- supervised classifier ablation, TTS-vs-natural over the BP KDE pool (364 natural / 57 synthetic speakers, NO MLAAD). Generated from exp43_clf_results.csv. Requires \usepackage{booktabs}. Balanced 5-fold CV (20 seeds x 5 folds, natural subsampled to synth size), XGBoost; mean $\pm$ SD. Rows sorted by EER (ascending).

% FakeBrAccent in-domain anti-spoofing (real vs fake), SPEAKER-DISJOINT CV. Generated from fakebr_indomain_results.csv. Requires \usepackage{booktabs}. Fakes are voice-conversion clones of the same speakers, so we use speaker-grouped 5-fold CV (a held-out speakers real and fake samples are both unseen), repeated over 10 seeds; pooled out-of-fold EER/ACC, mean $\pm$ SD. XGBoost on ZIPA-46 and SSL backbones (ECAPA, XLSR-PT, HuBERT, w2v-BERT). Rows sorted by EER.

A slightly different approach is required by FakeBRAccent, which does not have an official train/test split and contains samples from the same speaker in both spoof and natural parts, causing training/test leakage. Therefore, a speaker-disjoint cross-validation strategy is adopted that separates speakers by the dataset metadata --- otherwise, near-perfect accuracy is achieved with almost all feature groups, precisely because of leakage (the current baseline~\cite{fakebr_accent2025} does not address this issue). Results for this cross validation are presented in Table~\ref{tab:fakebr_indomain}.
\begin{table}[h]
\centering
\caption{Speaker-disjoint CV accuracy and EER (mean $\pm$ SD over 10 seeds of speaker-grouped 5-fold CV) on FakeBrAccent}
\label{tab:fakebr_indomain}
\begin{tabular}{lrcc}
\toprule
Feature set & Dim & ACC (\%) & EER (\%) \\
\midrule
\textbf{W2V2BERT+ZIPA} & \textbf{1070} & \textbf{91.7 $\pm$ 0.8} & \textbf{8.0 $\pm$ 0.8} \\
W2V2BERT & 1024 & 90.9 $\pm$ 0.7 & 9.0 $\pm$ 0.8 \\
ECAPA+ZIPA & 238 & 86.7 $\pm$ 0.6 & 13.4 $\pm$ 0.7 \\
XLS-R+ZIPA & 1070 & 85.1 $\pm$ 1.2 & 14.8 $\pm$ 1.3 \\
ZIPA & 46 & 85.2 $\pm$ 0.8 & 15.1 $\pm$ 1.0 \\
ECAPA & 192 & 83.6 $\pm$ 0.8 & 16.5 $\pm$ 0.7 \\
HuBERT+ZIPA & 1070 & 82.1 $\pm$ 0.9 & 18.5 $\pm$ 0.9 \\
XLS-R & 1024 & 78.4 $\pm$ 1.0 & 21.7 $\pm$ 1.3 \\
HuBERT & 1024 & 75.1 $\pm$ 0.7 & 25.2 $\pm$ 0.8 \\
\bottomrule
\end{tabular}
\end{table}

In both BRSpeechDF and FakeBRAccent evaluations, a drop of performance for our feature set is perceived (mostly when evaluated alone) when compared to the curated dataset experiments. This can be explained by the fact that these datasets comprise only voice-cloning samples, and not ``anonymized'' or general-purpose, mixed-identity voices such as the ones in the curated datasets and found in commercial and foundation open source TTS default voices. With voice cloning approaches, models can follow the accent realization choices of the target speaker, which narrows down their variability and can make the accent more consistent, thus making the job of accent disambiguation features more difficult. 

Yet another different approach was needed to evaluate detection on MLAAD-pt: it contains only synthetic speech, thus requiring training on other datasets. However, training on our customized dataset or even in BRSpeech resulted in near-perfect MLAAD detection for almost all approaches (such as Wav2Vec2Bert performs in~\cite{li2025where}), which may indicate that all out-of-domain data is being classified as spoofing. To avoid this phenomenon, it is necessary to add out-of-domain natural samples to the measurements in order to ensure generalization. 
\begin{table}[h]
  \centering
  \caption{EER(\%) under double leave-one-dataset-out.}
  \label{tab:exp52b_natlodo}
  \begin{tabular}{lrrrr}
  \toprule
  Feature set & Dim & Overall & Curated dataset & MLAAD \\
  \midrule
  \textbf{ECAPA+Formants} & \textbf{221}  & \textbf{6.09}  & \textbf{11.30} & \textbf{3.49} \\
  \textbf{ZIPA+PX+Formants}       & \textbf{90}  & \textbf{8.13}  & \textbf{12.13} & \textbf{6.13} \\
  XLS-R+ZIPA     & 1070 & 8.54  & 9.12  & 8.25 \\
  ECAPA+ZIPA    & 238  & 8.60  & 17.67 & 4.06 \\
  ECAPA         & 192  & 8.75  & 15.05 & 5.60 \\
  XLS-R+Z.+PX+Form.  & 1145 & 9.21  & 8.71  & 9.46 \\
  XLS-R          & 1024 & 10.57 & 9.74  & 10.98 \\
  W2V2BERT       & 1024 & 13.42 & 15.16 & 12.55 \\
  \bottomrule
  \end{tabular}
\end{table}
Thus, a double leave-one-dataset-out cross validation is proposed: each fold holds out one synthesis group and one natural corpus together, with the folds covering all synthetic--natural combinations. This simulates having all data available except for that system and that corpus, while keeping the natural side unseen as well. Results are presented in Table~\ref{tab:exp52b_natlodo}, where the EER is calculated for each held-out synthetic--natural pair, then averaged over all pairs and weighted by hold-out size, aggregated over dataset source (overall, curated, or MLAAD). Formant features combined with localized ECAPA-TDNN embeddings win here, and the explainable set demonstrates excellent generalization capabilities\footnote{Isolated per-system performance and full evaluation numbers are available at the companion page.} .

\section{Conclusions}
This work investigated whether the regional phonetic dilution introduced by modern Text-to-Speech can be turned into a mechanism for synthetic speech detection in Brazilian Portuguese. Building on localized phone recognition and classical signal processing over consonantal and vocalic realizations that carry high geographic variance, a set of lightweight, low-dimensional, speaker-level phonological profiles was built and evaluated on a curated accent-diverse pool and on consolidated anti-spoofing benchmarks.

Three findings stand out. First, the distributional gap between natural and synthetic speech over these accent-related features is large enough that an unsupervised Kernel Density Estimator adjusted only to natural speakers flags synthetic voices as out-of-distribution, establishing dialectal inconsistency as a separable and interpretable cue rather than an artifact of an overfitted classifier, while still capturing natural samples as being within the distribution (74.3\% AUC before t-SNE transformations, 93.9\% after). Second, this separability is interpretable: synthetic voices exhibit shifted distributions, distorted spectral shapes, and distinguishable vowel spaces, together with measurably higher within-utterance marker switching. With the distributions being statistically different, they align with the hypothesis of models collapsing distinct regional variants into less coherent realizations. Third, these features are also useful for spoofing detection in practice. In BRSpeechDF, they help the best configuration reach \mbox{2.00\%} EER. In FakeBrAccent and MLAAD-pt, they match prior baselines in uncontrolled domain tests with high accuracy, while showing strong cross-dataset generalization (15\% and 6.3\% EER, respectively) within stricter dataset-isolated validation strategies.

\section*{Acknowledgments}
This work was partially funded by the Brazilian Federal Agency for Support and Evaluation of Graduate Education, CAPES (001), and the Brazilian National Council for Scientific and Technological Development, CNPq (grants 409179/2024-8 and 306395/2025-8 --- National Institute of Science and Technology (INCT) STREAM).

\section*{AI-generated content disclosure}
Anthropic's AI models (i.e. Sonnet 4.6,  Opus 4.8, and Opus 5) were employed in this work to help with text, tables, figures, and experimental code, under strict human supervision from the authors. More specifically, grammar and spelling were double checked and improved, all tables and figures were created by AI-generated code that extracted tabular data obtained in the experiments, and the source code for the experiments was partially generated and auto-completed by the AI.

% This work was partially supported by XXXXXXX (XX/XXXXX-X).

\bibliographystyle{ieeetr}
\bibliography{references}

\end{document}